\documentclass[a4paper,11pt]{article} 
\usepackage{pos}
\title{The Monitoring, Logging, and Alarm  system for the Cherenkov Telescope Array}
\ShortTitle{MON \& AAS  systems for CTA}
\manuallySeparateAuthors
\author*[a]{Alessandro Costa}
\author[a]{, Kevin Munari}
\author[a]{, Federico Incardona}
\author[a]{, Pietro Bruno}
\author[b]{, Stefano Germani}
\author[a]{, Alessandro Grillo}
\author[c]{, Igor Oya}
\author[a]{, Eva Sciacca}
\author[a]{, Ugo Becciani}
\author[a]{and Mario Raciti}
\author{, for the CTA Consortium}
\affiliation[a]{INAF, Osservatorio Astrofisico di Catania, Via S Sofia 78, I-95123 Catania, ITALY}
\affiliation[b]{Universit\`a di Perugia, Dipartimento di Fisica e Geologia, IT}
\affiliation[c]{CTA Observatory gGmbH}
\affiliation[d]{INAF, Osservatorio di Astrofisica e Scienza dello Spazio di Bologna, IT}
\emailAdd{alessandro.costa@inaf.it}
\abstract{We present the current development of the Monitoring, Logging and Alarm subsystems in the framework of the Array Control and Data Acquisition System (ACADA) for the Cherenkov Telescope Array (CTA). The Monitoring System (MON) is the subsystem responsible for monitoring and logging the overall array (at each of the CTA sites) through the acquisition of monitoring and logging information from the array elements. The MON allows us to perform a systematic approach to fault detection and diagnosis supporting corrective and predictive maintenance to minimize the downtime of the system. We present a unified tool for monitoring data items from the telescopes and other devices deployed at the CTA array sites. Data are immediately available for the operator interface and quick-look quality checks and stored for later detailed inspection.
The Array Alarm System (AAS) is the subsystem that provides the service that gathers, filters, exposes, and persists alarms raised by both the ACADA processes and the array elements supervised by the ACADA system. It collects alarms from the telescopes, the array calibration, the environmental monitoring instruments and the ACADA systems. The AAS sub-system also creates new alarms based on the analysis and correlation of the system software logs and the status of the system hardware providing the filter mechanisms for all the alarms. Data from the alarm system are then sent to the operator via the human-machine interface.}
\FullConference{37$^{\rm{th}}$ International Cosmic Ray Conference (ICRC 2021)\\
		July 12th -- 23rd, 2021\\
		Online -- Berlin, Germany}
\begin{document} 
\maketitle
%
%
\section{Introduction and motivation}
The Cherenkov Telescope Array (CTA)\cite{2019scta.book.....C, 2013APh....43....3A} will be the largest and most advanced ground-based facility for detection of very-high-energy electromagnetic radiation, from 20 GeV to 300 TeV. When entering the atmosphere, this radiation generates secondary charged particle cascades that can be detected directly or, as in the case of CTA, through the Cherenkov radiation they emit. Since the area hit by this light is wide, in the order of $10^5\,\mathrm{m^2}$, multiple telescopes are required to intercept it all, and to reconstruct the properties (energy, direction) of the primary gamma-ray who generated the cascades. CTA will be composed of tens of telescopes deployed at the Northern and Southern Hemispheres to achieve full-sky coverage, and an angular resolution of about 1 arcminute at high energies. Typical phenomena that can be investigated include supernovae, supernova remnants, pulsars and pulsar wind nebulae, binary stellar systems, interacting stellar winds, various types of active galaxies, gamma-ray bursts, and gravitational wave transients. By means of its observation, CTA is expected to shed light on some unresolved astrophysics questions such as the role of relativistic cosmic particles on star formation and galaxy evolution, the physics in the proximity of neutron stars and black holes, or the nature of the dark matter. Together with the scientific data produced by CTA, a big volume of housekeeping and auxiliary data coming from weather stations, instrumental sensors, logging files, etc., must be collected as well. In order to ingest the whole amount of data coming from tens of telescopes, a complex software architecture is required that must be able to face such a cutting-edge technological challenge.\\
Quality requirements such as reliability, performance, scalability, availability are of fundamental importance. the Array Control and Data Acquisition System (ACADA) addresses these requirements, and supervises the data taking and the telescope control. It also provides the user interface for the site operators and astronomers.\\ 
In this paper we describe the Monitoring System (MON) and the Array Alarm System (AAS) which are part of ACADA. The MON is responsible for monitoring and logging the overall array system through the acquisition of monitoring and logging points from the elements of the array, and for making these data immediately available for the operator interface and for quick-look quality checks, as well as to store them for later detailed inspection. The AAS is responsible for collecting alarms from telescopes, array calibration and environmental monitoring instruments, and the ACADA systems itself. The AAS creates also new alarms based on the analysis and correlation of the system software logs and status of the system hardware and provides filter mechanisms for all alarms. The data from the alarms are sent to the operator via the human-machine interface (HMI). We previously presented the MON and AAS systems in \cite{10.1117/12.2560697}. In this paper we describe the whole MON and AAS system as well the technological choices for the implementation.\\
The paper is organized as follows. Section 2 reviews the basics of IoT and NoSQL technologies; Section 3 introduces the architecture of the monitoring, logging and alarm systems; Section 4 introduces the technologies involved in the development; Section 5 presents the conclusions and the future perspectives.
\section{Background and basic concepts}
In this section we recall the basics of IoT and NoSQL technologies and overview the characteristics essential for their integration.
\subsection{Background of Big Data and IoT}\label{big data}
The term, IoT was used to describe a system where the Internet is connected to the physical world via ubiquitous sensors. IoT involves by definition a large amount \cite{aguzzi2013definition} of information sources (i.e., the things), producing a huge amount of semi-structured data \cite{buneman1997semistructured} which also have the three characteristics typical of Big Data: volume \cite{zikopoulos2011understanding} (i.e., data size), variety (i.e., data types), and velocity (i.e., data generation frequency).\\
Big data is a term utilized to refer to the increase in the volume of data that are difficult to store, process, and analyze through traditional database technologies.
Being the term “big data” relatively new in IT and business it has been defined several times in literature and  with slightly different meanings. For instance \cite{cox1997managing} referred to big data as a large volume of scientific data for visualization; \cite{manyika2011big} defined big data as the amount of data just beyond technology's capability to store, manage, and process efficiently.\\Meanwhile, \cite{zikopoulos2013harness} defined big data as characterized by four Vs: \textbf{volume}, \textbf{variety}, \textbf{velocity} and \textbf{value}. 
Volume refers to the amount of all types of data generated from different sources and continue to expand; Variety refers to the different types: since a common big data scenario collects different information like video, text, pdf, and graphics on social media, as well as  technical data from sensors. Velocity refers to the speed of data transfer. Value is the most important aspect of big data, it refers to the process of discovering hidden values from large datasets leveraging various types of data analysis.
\subsection{CTA ACADA monitoring and logging data characterization}\label{ACADA data characterization}
In  the  framework  of CTA ACADA we expect about 200.000 monitoring points sampled between 1 and 5 Hz for a maximum data rate for writing operations of 26 Mbps for the monitoring system including the alarms. A maximum  rate of about 1 Gbps has been estimated for storing log information. We are considering here the characteristics of our data compared with the definition on \textit{big data} that was given in section \ref{big data}.
An high volume and rate of data can here be easily identified: volume and velocity attributes describe therefore our data-set. The data structure for monitoring and logging data in ACADA has been specified instead with a well structured data scheme \cite{Avro}. No variety attribute can be identified. As of ``value'', the logging and monitoring system aim at providing a solid framework for identifying occurred problems and for implementing  predictive maintenance techniques. The more those data are known and mined the less the whole system will be affected by interruptions. CTA ACADA monitoring and logging data-set can definitively claim the title of ``big data'' being described by three out of four of the attributes that can identify such a kind of data.
\subsection{NoSQL technologies in comparison}
NoSQL databases were born in the big data era to deal with high-volume and variety of data traveling very fast. It is possible to classify NoSQL database according to their storage type. Some relevant examples are Couchbase \cite{Couchbase} or Redis \cite{Redis}, which are essentially in-memory databases, or column store database such as Cassandra \cite{Cassandra} and Hbase \cite{Hbase}, or document store databases as CouchDB \cite{CouchDB}, MongoDB \cite{MongoDB}, Elasticsearch \cite{Elasticsearch}, and many others. In spite of the classification, every distributed database must obey to the \textit{CAP theorem}, which asserts that, in the presence of a network failure, the system has to choose between consistency and availability. The first being the ability of the system to show to all the clients the same data at the same time, while the second is the ability to correctly respond to every client request. So that, different technologies follow different approaches to the CAP theorem providing the user with different solutions. For instance, in presence of network partition, Cassandra and CouchDB guarantee availability, while MongoDB, Redis and Hbase guarantee consistency \cite{Lourenco15}. Comparing some qualitative attributes of the most popular NoSQL databases, it results that Cassandra ensures at same time great write-performance, scalability, availability and consistency \cite{Lourenco15}, in contrast to MongoDB that instead is by far the most popular \cite{DBEngine}. As already mentionend, the main difference between these two systems is the storage type. Besides, Cassandra exploits a peer-to-peer architecture with no hierarchy among nodes. On the contrary, MongoDB uses a master/slave approach and sharding to spread load. Strong differences exist in terms of performance between the twos. In fact, with respect to MongoDB, Cassandra is optimized to work with larger volumes of data \cite{Abramova2013}. Furthermore, while MongoDB is faster than Cassandra for what concern readings, things revert for writing operations. For this reason, Cassandra appears to be the best solution for large critical sensor applications \cite{VDVVDWM12}. Being the data produced by the the monitoring and logging system characterized by an high volume and rate in writing operations as was demonstrated in \ref{ACADA data characterization}, Cassandra was then the most suitable and solid database  solution for our purposes.
\section{Logging, Monitoring and Alarm Systems Architecture}
\begin{figure}
    \centering
    \includegraphics[width=1\textwidth]{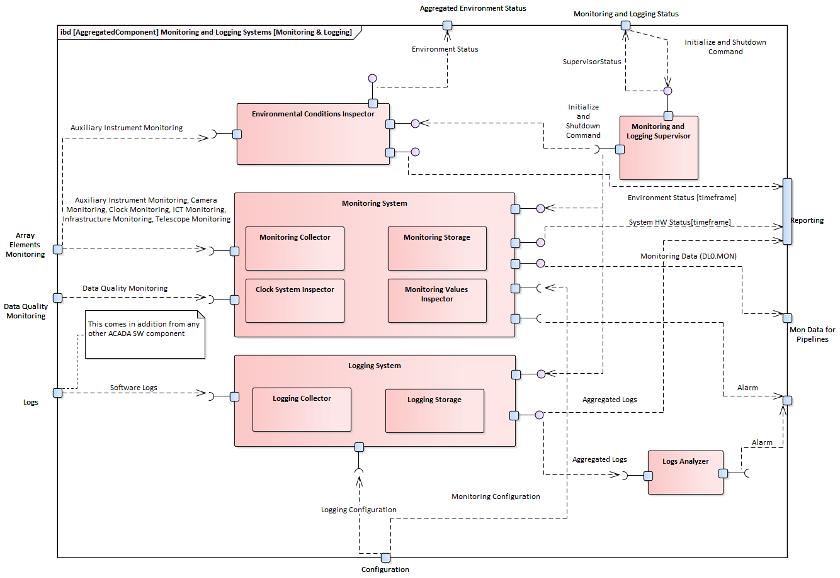}\hfill
    \includegraphics[width=0.6\textwidth]{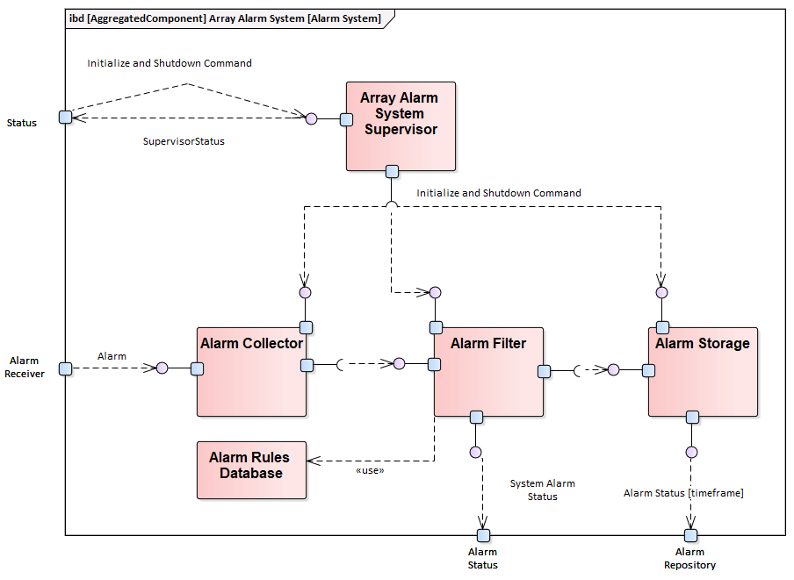}
    \caption{Software architecture for the Monitoring, Logging Systems (Up); Alarm System (Down) }
    \label{fig_1}
\end{figure}

The Monitoring and Logging subsystems (hereafter MON) provide  services for monitoring data items from the Telescopes and other devices deployed at the CTA array sites and making those data immediately available for the operator interface and for quick-look quality checks, as well as to store them for later detailed inspection. This architecture takes advantage of continuous technological evolution \cite{costa:icalepcs2019-mopha032} to respond to the challenges posed by the operation of the array, in particular to satisfy the reliability. MON includes the production of the software for the Monitoring System Logging System and Logs Analyser.\\
MON  is composed of the following main building blocks: (Fig \ref{fig_1})\\
\textbf{Monitoring System}: Provides the services that gather monitoring data (i.e. time-series data from instruments sensors and statuses at typically ~1 Hz rates) from the Telescopes and other instruments, such as environment monitoring devices, and stores them in a local database. The monitoring system  works continuously to record any monitoring data made availale by Array Elements, which also includes the data points required for engineering purposes.\\
\textbf{Monitoring and Logging Supervisor}: It receives  startup and  shut-down commands by the ACADA RM (Resource Manager) and passes 	them to the MON systems. It provides its status information to the RM.\\
\textbf{Environmental Conditions Inspector}: A component that accumulates the data from the central calibration devices to produce indicators for the status of the environment. Most of the environmental monitoring data is received directly from the monitoring system. The component processes the monitoring data, puts associated data elements together, and determines and stores the status of the environment.\\
\textbf{Monitoring Value Inspector}:  A component that provides the capability to re-sample the monitoring information coming in the form of irregular and unevenly spaced time series data to a consistent and regular frequency. The component processes the monitoring events and raises alarms if data is above the threshold for a predefined period of time.\\

The Logging System is composed of the following main building blocks: (Fig \ref{fig_1})\\
\textbf{Logging System}:  Gets logging information from relevant software components and stores it. This logging comes in three flavours: software logs provided by elements using the control framework, software logs of the observation scripts, software logs produced by low-level firmware, that require reformatting to adapt to the rest of the logs\\
\textbf{Logs Analyzer}:  A tool to analyse logging data information to trigger further alarms, as well as warnings for the technical crew.\\

The AAS is composed of the following main building blocks: (Fig. \ref{fig_1}) \\
\textbf{Alarm Collector}: Provides the services to collect any alarm raised by the Array Elements or ACADA components.\\
\textbf{Alarm Filter}: Provides means to filter, merge and reduce alarms according to defined rules.\\
\textbf{Alarm Rules Database}: A database defining the alarm reduction rules for the Alarm Filter.\\
\textbf{Alarm Storage}:  Local repository to store alarms and reactions to alarm history.\\
\textbf{AAS Supervisor}: Manager component for the AAS, connected with to the supervision tree provided by the RM.
\section{Technologies}
According to what has been decided for both CTA sites installations, our MON and AAS systems are integrated with the ALMA Common Software (ACS) \cite{ACS}, an open-source framework on which the software operating the ALMA observatory is based on. ACS allows the use of the following programming languages: C++, Java, and Python. Although there is more C++ and Python tradition in CTA, we opted for Java because ACS is mostly implemented in Java and a lot of ACS documentation is available for Java developers. Furthermore, the robustness of Java is considered more important than the extra performance gain with C++. Java is, in fact, robust, ease of use, platform-independent, and secure. ACS makes use of Apache Maven \cite{Maven} and we also take advantage of it to build and manage in an automated way our Java-based systems. As previously stated, during the activities of the array site, MON is designed to acquire monitoring and logging points of ACADA. More specifically, MON can access and monitor ACS and OPC-UA data sources. To this aim, MON makes use of Eclipse Milo SDK \cite{Milo}, which provides a pure-Java, open-source implementation of the OPC-UA 1.03 client and server specifications. To exchange the acquired data among the heterogeneous ACADA subsystems, we opted for Apache Avro \cite{Avro}, a data serialization framework that uses JSON forhttps://www.overleaf.com/project/60cb6c4ca023bc0f99748172 defining schemas the information exchanged must be compliant with. Finally, to support such a volume of data, we make use of Apache Kafka \cite{Kafka}, a distributed event streaming platform designed to handle data streams from multiple sources and deliver them to multiple consumers. Besides, to forward and centralize logs generated by ACADA, we use a set of distributed lightweight shippers based on Elastic Filebeat \cite{Filebeat}. Those log events are ingested, filtered and manipulated by a centralized log aggregator based on Elastic Logstash \cite{Logstash}, which acts as a data processing pipeline that, in the end, sends them to Apache Kafka. As previously stated, we opted for Apache Cassandra as our database management system (DBMS), which is specifically designed to handle large amounts of data. Finally, we make use of the Docker platform \cite{Docker} to easily distribute, replicate and scale our deployment environment, packaging the technologies described above in containers.
\section{Conclusion and Future Directions}
We presented the architecture of a system that monitors and logs the data needed to improve the operational activities of a large scale telescope array. The system was designed and built considering the current software
tools and concepts coming from Big Data and Internet of Things. The software stack is based on open source software, thus reducing the need for unnecessary extra software development.
Future work is planned to integrate Machine Learning algorithms to perform anomaly detection, failure prediction and  to manage complex events \cite{rao2012cloud}.
%
%
\bibliographystyle{JHEP}
\bibliography{report} 

\providecommand{\href}[2]{#2}\begingroup\raggedright\begin{thebibliography}{10}

\bibitem{2019scta.book.....C}
\emph{{Science with the Cherenkov Telescope Array}} (2019),
  \href{https://doi.org/10.1142/10986}{10.1142/10986}.

\bibitem{2013APh....43....3A}
\emph{{Introducing the CTA concept}},
  \href{https://doi.org/10.1016/j.astropartphys.2013.01.007}{\emph{Astroparticle
  Physics} {\bfseries 43} (2013) 3}.

\bibitem{10.1117/12.2560697}
A.~Costa, G.~Tosti, J.~Schwarz, P.~Bruno, A.~Bulgarelli, A.~Calanducci et~al.,
  \emph{{Architectural design and prototype for the logging, monitoring, and
  alarm system for the ASTRI mini-array}},  in \emph{Software and
  Cyberinfrastructure for Astronomy VI}, J.C.~Guzman and J.~Ibsen, eds.,
  vol.~11452, International Society for Optics and Photonics, SPIE, 2020,
  \href{https://doi.org/10.1117/12.2560697}{DOI}.

\bibitem{aguzzi2013definition}
S.~Aguzzi, D.~Bradshaw, M.~Canning, M.~Cansfield, P.~Carter, G.~Cattaneo
  et~al., \emph{Definition of a research and innovation policy leveraging cloud
  computing and iot combination}, {\emph{Final Report, European Commission,
  SMART} {\bfseries 37} (2013) 2013}.

\bibitem{buneman1997semistructured}
P.~Buneman, \emph{Semistructured data},  in \emph{Proceedings of the sixteenth
  ACM SIGACT-SIGMOD-SIGART symposium on Principles of database systems},
  pp.~117--121, 1997.

\bibitem{zikopoulos2011understanding}
P.~Zikopoulos and C.~Eaton, \emph{Understanding big data: Analytics for
  enterprise class hadoop and streaming data}, McGraw-Hill Osborne Media
  (2011).

\bibitem{cox1997managing}
M.~Cox and D.~Ellsworth, \emph{Managing big data for scientific visualization},
   in \emph{ACM siggraph}, vol.~97, pp.~21--38, 1997.

\bibitem{manyika2011big}
J.~Manyika, M.~Chui, B.~Brown, J.~Bughin, R.~Dobbs, C.~Roxburgh et~al.,
  \emph{Big data: The next frontier for innovation, competition, and
  productivity}, McKinsey Global Institute (2011).

\bibitem{zikopoulos2013harness}
P.C.~Zikopoulos, D.~Deroos and K.~Parasuraman, \emph{Harness the power of big
  data: The IBM big data platform}, McGraw-Hill, (2013).

\bibitem{Avro}
``Apache avro.'' \url{https://avro.apache.org/}.

\bibitem{Couchbase}
``Couchbase.'' \url{https://www.couchbase.com}.

\bibitem{Redis}
``Redis.'' \url{https://redis.io}.

\bibitem{Cassandra}
``Apache cassandra.'' \url{https://cassandra.apache.org}.

\bibitem{Hbase}
``Apache hbase.'' \url{https://hbase.apache.org}.

\bibitem{CouchDB}
``Apache couchdb.'' \url{https://couchdb.apache.org}.

\bibitem{MongoDB}
``Mongodb.'' \url{https://www.mongodb.com}.

\bibitem{Elasticsearch}
``Elasticsearch.'' \url{https://www.elastic.co}.

\bibitem{Lourenco15}
J.R.~Lourenço, B.~Cabral, P.~Carreiro, M.~Vieira and J.~Bernardino,
  \emph{Choosing the right nosql database for the job: a quality attribute
  evaluation}, \href{https://doi.org/10.1186/s40537-015-0025-0}{\emph{Journal
  of Big Data} {\bfseries 2} (2015) }.

\bibitem{DBEngine}
``Db-engines ranking.'' \url{https://db-engines.com/en/ranking}.

\bibitem{Abramova2013}
V.~Abramova and J.~Bernardino, \emph{Nosql databases: Mongodb vs cassandra},
  {\emph{C3S2E '13: Proceedings of the International C* Conference on Computer
  Science and Software Engineering} (2013) }.

\bibitem{VDVVDWM12}
J.S.V.D.~Veen, B.V.D.~Waaij and R.J.~Meijer, \emph{Sensor data storage
  performance: Sql or nosql, physical or virtual},  pp.~431--438, 2012,
  \href{https://doi.org/10.1109/CLOUD.2012.18}{DOI}.

\bibitem{costa:icalepcs2019-mopha032}
A.~Costa et~al., \emph{{Big Data Architectures for Logging and Monitoring Large
  Scale Telescope Arrays}},  in \emph{Proc. ICALEPCS'19}, no.~17 in
  International Conference on Accelerator and Large Experimental Physics
  Control Systems, pp.~268--271, JACoW Publishing, Geneva, Switzerland, 08,
  2020, \href{https://doi.org/10.18429/JACoW-ICALEPCS2019-MOPHA032}{DOI}.

\bibitem{ACS}
``Alma common software.'' \url{https://www.eso.org/projects/alma/develop/acs/}.

\bibitem{Maven}
``Apache maven.'' \url{https://maven.apache.org/}.

\bibitem{Milo}
``Eclipse milo.'' \url{https://projects.eclipse.org/projects/iot.milo}.

\bibitem{Kafka}
``Apache kafka.'' \url{https://kafka.apache.org/}.

\bibitem{Filebeat}
``Elastic filebeat.'' \url{https://www.elastic.co/beats/filebeat}.

\bibitem{Logstash}
``Elastic logstash.'' \url{https://www.elastic.co/logstash}.

\bibitem{Docker}
``Docker.'' \url{https://www.docker.com/}.

\bibitem{rao2012cloud}
B.P.~Rao, P.~Saluia, N.~Sharma, A.~Mittal and S.V.~Sharma, \emph{Cloud
  computing for internet of things \& sensing based applications},  in
  \emph{2012 Sixth International Conference on Sensing Technology (ICST)},
  pp.~374--380, IEEE, 2012.

\end{thebibliography}\endgroup
\end{document}